\documentclass[conference]{IEEEtran}
\usepackage{cite}

\ifCLASSINFOpdf
  \usepackage[pdftex]{graphicx}
\else
  \usepackage[dvips]{graphicx}
\fi
\usepackage{amsmath}
\usepackage{amssymb}

\DeclareMathOperator*{\argmax}{arg\max}
\usepackage{subfigure}
\begin{document}
%
\title{\huge{Physical Layer Authentication for Non-coherent Massive SIMO-Based Industrial IoT Communications}}
\author{\IEEEauthorblockN{Zhifang Gu\IEEEauthorrefmark{1}, He Chen\IEEEauthorrefmark{2}, Pingping Xu\IEEEauthorrefmark{1}, Yonghui Li\IEEEauthorrefmark{3}, and Branka Vucetic\IEEEauthorrefmark{3}}
                \IEEEauthorblockA{\IEEEauthorrefmark{1}National Mobile Communications Research Laboratory, Southeast University, Nanjing, China}
                \IEEEauthorblockA{\IEEEauthorrefmark{2}Department of Information Engineering, The Chinese University of Hong Kong, Hong Kong, China}
                \IEEEauthorblockA{\IEEEauthorrefmark{3}School of Electrical and Information Engineering, The University of Sydney, Sydney, Australia \\
                  \IEEEauthorrefmark{1}\{zhifang\_gu, xpp\}@seu.edu.cn,
                  \IEEEauthorrefmark{2}he.chen@ie.cuhk.edu.hk,
                  \IEEEauthorrefmark{3}\{yonghui.li, branka.vucetic\}@sydney.edu.au}
 }

%


\maketitle

\begin{abstract}
Achieving ultra-reliable, low-latency and secure communications is essential for realizing the industrial Internet of Things (IIoT). Non-coherent massive multiple-input multiple-output (MIMO) has recently been proposed as a promising methodology to fulfill ultra-reliable and low-latency requirements. In addition, physical layer authentication (PLA) technology is particularly suitable for IIoT communications thanks to its low-latency attribute. A PLA method for non-coherent massive single-input multiple-output (SIMO) IIoT communication systems is proposed in this paper. Specifically, we first determine the optimal embedding of the authentication information (tag) in the message information. We then optimize the power allocation between message and tag signal to characterize the trade-off between message and tag error performance. Numerical results show that the proposed PLA is more accurate then traditional methods adopting the uniform tag when the communication reliability remains at the same level. The proposed PLA method can be effectively applied to the non-coherent system. 
\end{abstract}


%
\IEEEpeerreviewmaketitle

\section{Introduction}
The fast development of the Internet of Things (IoT) has promoted innovations in many fields. The application of the IoT in the industrial sector, referred to as industrial IoT (IIoT), has recently attracted tremendous attention from researchers and engineers owing to its ability to improve the efficiency and productivity of industry. Compared with traditional industrial networks mainly based on wired cables, wireless communications are more suitable for the IIoT due to low maintenance expenditure, flexible deployment, and higher long-term reliability \cite{a2}. However, ultra-reliable, low-latency and secure requirements of the IIoT represent main challenges for wireless design \cite{a3} \cite{a4}. Wireless channels suffer from path-loss, shadowing, fading and interference, thus it is very challenging to design wireless networks to achieve the ultra-reliable transmission\cite{a5}. Moreover, the broadcast characteristic of wireless channels makes the IIoT systems more vulnerable to attacks \cite{a6}. Non-coherent massive multiple-input multiple-output (MIMO) has recently been proposed as a promising methodology to meet ultra-reliable and low-latency requirements of IIoT communications \cite{a7}, which uses multiple receive antennas to reduce the effects of fading and uncorrelated noise in wireless channels and boost system reliability. Besides, non-coherent massive MIMO uses energy-based modulation to achieve low latency by avoiding channel estimation and by applying fast non-coherent detection \cite{a8} \cite{a9}. Two security services have been commonly considered in the IIoT, including integrity and authenticity, which are essential in IIoT systems. Message authentication code (MAC) is a prevalent mechanism to provide these two services. Conventional systems realize message authentication by attaching a MAC to the message and this authentication process is completed above the physical layer \cite{aa1}, e.g., transport layer security (TLS) protocol in the transport layer and Wi-Fi protected access II (WPA2) protocol in the network layer. However, these conventional mechanisms may not be able to meet the stringent low-latency requirement of IIoT communications. Because short packet transmission is one of the characteristics of the IIoT, the transmission overhead for the MAC can be large and excessive in the short packet transmission with small payload, occurring relatively large delay. In addition, the authentication process can only be completed after the data has been transferred to upper layers, which leads to low efficiency. Therefore, we aim to propose a message authentication method at the physical layer which can also reduce transmission overhead.

Physical layer security, according to its implementation method, can be divided into two categories. The first category is based on the information-theoretic approach, which was proposed by Shannon \cite{aa3} and further developed by Wyner with the wiretap channel model \cite{aa4}. This kind of approach only guarantees data confidentiality by preventing eavesdropper from understanding the information, but other security services like data integrity and authenticity are not considered. The second category is based on the signal and channel features, which aims to provide authenticity and data integrity. In the IIoT scenario, active attacks (e.g., modification, masquerade or replay attack) are much more harmful than passive attacks (e.g., eavesdropping)\cite{aa6}. Therefore, we consider physical layer authentication (PLA) from the perspective of the second category. Existing PLA methods have two forms: passive and active \cite{a11}. Passive PLA utilizes the intrinsic features of communication systems to authenticate the transmitter, such as radio signal strength indicator, channel state information (CSI) and radio frequency fingerprints \cite{a6}. These features were thoroughly analyzed in \cite{a12} with a theoretical model and experimental validation. The results of \cite{a12} revealed that the intrinsic features are not reliable in practical scenarios due to the device mobility, wireless fading channels and indistinguishable RF fingerprints. In contrast, active PLA refers to the methods in which the transmitter sends additional information (normally referred to as tag) for authentication at the physical layer. Active PLA features embedding a tag in the message information and does not take extra time to transmit the tag. Thanks to its potential to meet the low-latency requirement, active PLA has advantages over conventional authentication methods sending a message and its tag separately \cite{a14}.

The key issue of implementing active PLA is how to embed a tag in message information at the physical layer. Several methods dealing with this issue have been published. The tag was added as noise in \cite{a15}: different additional angle offsets to normal quadrature phase shift keying (QPSK) indicate different tag bits. 4/16 hierarchical quadrature amplitude modulation (QAM) was applied in \cite{a16} to transmit a message and its tag simultaneously, where a 4-QAM tag constellation is superimposed on a 4-QAM message constellation. Challenge response PLA was introduced in \cite{a17} and the authentication information was embedded during a ``challenge and response" process. Although these active PLA methods transmit a tag and a message at the same time, they all need to send separate pilots for channel estimation to acquire the instantaneous CSI. In \cite{a18}, a tag is embedded in the original pilot to form a new pilot, and the tag detection is completed by a correlation operation. Then the new pilot signal is used to estimate CSI for message recovery. Note that these existing active PLA methods are not suitable for non-coherent massive MIMO-based IIoT systems, in which no estimation of the instantaneous CSI is performed. To our best knowledge, how to perform active PLA for non-coherent systems is still an open problem.

As the first attempt to fill this gap, we focus on designing an active PLA mechanism for non-coherent massive single-input multiple-output (SIMO) IIoT systems. As elaborated in \cite{a19}, non-negative pulse amplitude modulation (PAM) is a favorable scheme for the considered system. In the system, the variance of the received signal power increases as the transmitted signal amplitude increases. In this context, the tag embedding constellation pattern is not necessary to be uniform as in the existing methods, but needs to be optimized according to the message constellation. The tag embedding design becomes a nontrivial problem as the increased tag signal power reduces the error rate of tag while increasing the error rate of message, leading to the error performance trade-off between message and tag. In this paper, we manage to find an optimal 1-bit tag embedding design based on a given message constellation. Then for a fixed average system power, we attain the optimal power allocation of message and tag signals to characterize the trade-off between message and tag error performance, which can provide useful insights for practical system design. 
\section{System Model}\label{SystemModelSec}
\begin{figure}[tbp]
	\centering
	\includegraphics[width=0.45\textwidth]{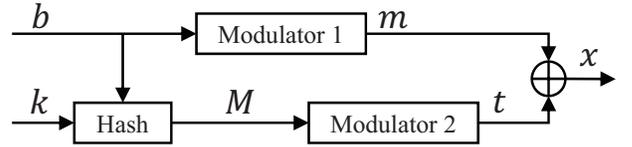}
	\caption{Diagram of the PLA at the transmitter side.}
	\label{Original Transmitter}
	\vspace{-1.5 em}
\end{figure}

In this paper, we consider the PLA in a massive SIMO-based IIoT communication system, where multiple sensors transmit data to a controller with $N$ antennas. Each sensor has one single antenna and these sensors send data to the controller with time-division multiple access (TDMA) manner. An attacker is within the range of this wireless communication system, who can receive the signal from sensors and send malicious signals to the controller. To meet the low-latency requirement of the IIoT, we adopt non-negative PAM at the transmitter and non-coherent maximum likelihood (ML) detector at the receiver \cite{a7,a8,a9}. Since only statistics of the channel are needed in this method, the channel estimation process is not required and the authentication can be executed faster. To realize PLA, the transmitter embeds the tag signal in the message signal at the physical layer, as illustrated in Fig.~\ref{Original Transmitter}. MAC is denoted by $M$, which is generated by message bits $b$ and secret key $k$ with a hash function, and it is given by
\begin{equation}
M={\rm{hash}}(b,k).
\end{equation}
The modulated signal of $b$ and $M$ are denoted by $m$ and $t$, which are termed message signal and tag signal, respectively. The transmitted signal $x$ is then determined by
\begin{equation}
x=\sqrt{|m|^2+|t|^2},
\end{equation}
where $m\in \mathcal{M}=\{m_i|i=1, \cdots , L_m\}$ and $t\in\mathcal{T}=\{t_{i,j}|i=1, \cdots , L_m;\ j=1, \cdots, L_t\}$. $L_m$ and $L_t$ are the number of constellation points of message signal and tag signal, respectively. The transmitted signal that involves message signal $m_i$ is denoted as $x_i$. The average power of message signal $E_m$ and the average power of tag signal $E_t$ are constrained by the total average system power $E_{tot}$
\begin{equation}
E_m+E_t\le E_{tot},
\end{equation}
where $E_m=\frac{1}{L_m}\sum \limits_{i=1}^{L_m}|m_i|^2$ and $E_t=\frac{1}{L_mL_t}\sum \limits_{i=1}^{L_m}\sum \limits_{j=1}^{L_t}|t_{i,j}|^2$.
The received signal $\bf{y}$ at the controller in the considered massive SIMO system can be represented by
\begin{equation}
{\bf{y}}={\bf{h}}x+{\bf{n}},
\label{y}
\end{equation}
where ${\bf{h}}=[h_1, \cdots, h_N]^T$ is the SIMO channel vector and ${\bf{n}}=[n_1, \cdots, n_N]^T$ is the noise vector between the sensor and controller. We assume that $\bf{h}$ is a circularly symmetric complex Gaussian random vector, specifically, each element of $\bf{h}$ is independently and identically distributed with zero mean and unit variance (i.e., Rayleigh fading). Another assumption is that $\bf{n}$ is a circularly symmetric complex Gaussian random vector which is independent of $\bf{h}$. The mean vector of  $\bf{n}$  is a zero vector and its covariance matrix is $\sigma^2{\bf I}_N$ . $\sigma^2$ is assumed to be known and ${\bf I}_N$ is an $N$-dimensional unit matrix.

Two steps are required in the signal detection at the receiver side. First, following the ML detection rule, the estimated message signal $\hat{m}$ is obtained by solving the problem
\begin{equation}
{\hat{m}}=\argmax_{m\in\mathcal{M}} {f({\bf{y}}|m)},
\label{ML1}
\end{equation}
where $f({\bf{y}}|m)$ is the probability density function (PDF) of $\bf{y}$ conditioned on $m$. Second, the estimated tag signal $\hat{t}$ is obtained based on $\hat{m}$ with the ML rule,
\begin{equation}
{\hat{t}}=\argmax_{t\in\mathcal{T}} {f({\bf{y}}|\hat{m},t)},
\end{equation}
where $f({\bf{y}}|\hat{m},t)$ is the PDF of $\bf{y}$ conditioned on $\hat{m}$ and $t$.

After the detection, we can get the estimated message bits $b'$ and the estimated MAC $M'$ with the demodulation of $\hat{m}$ and $\hat{t}$, respectively. Then a new MAC $M_n$ is calculated with $b'$ and $k$ by the same hash function, i.e., $M_n={\rm hash}(b',k)$. The authentication process is completed by comparing $M'$ with $M_n$. If $M'$ and $M_n$ are identical, the message can be regraded as coming from a legitimate user\footnote{Note that time stamps or session identifiers are required to resist replay attacks.} and not being tampered with. Otherwise, this message will be discarded because it is not authenticated successfully.  
\section{Preliminaries and Detection Rules}\label{SignalDesignSec}
In this section, we first introduce preliminaries on non-negative PAM design for message constellation in the massive SIMO system, and then present the 1-bit tag embedding design when the message constellation points are given.
\subsection{Preliminaries on Message Constellation Design}\label{pre}
\begin{figure}[bp]
\vspace{-1.5em}
\centering
\subfigure[Message constellation design and decision thresholds]{
\label{Constellation_a}
\includegraphics[width=0.46\textwidth]{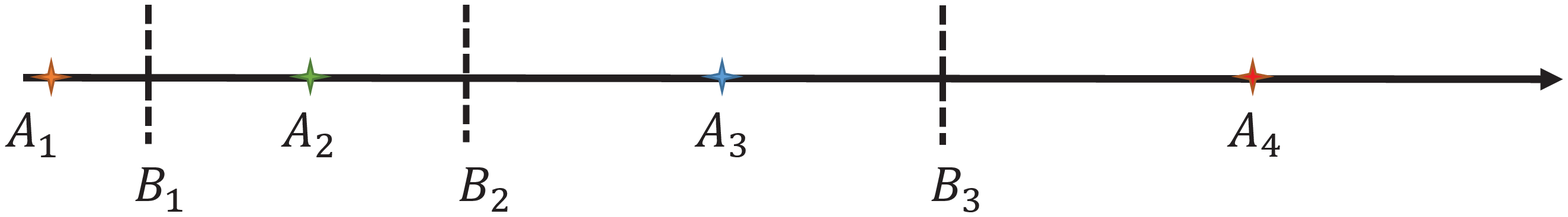}}
\subfigure[Tag Embedding design and updated decision thresholds.]{
\label{Constellation}
\includegraphics[width=0.46\textwidth]{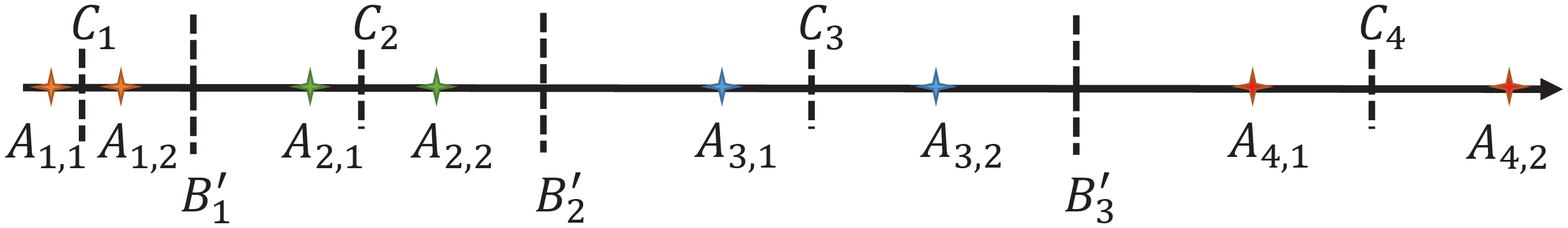}}
\caption{Signal design and decision thresholds.}
\label{Fig.main}
\end{figure}

We now consider only a message signal $m\in\mathcal{M}$ is transmitted. The problem \eqref{ML1} has been solved in \cite{a9}, which showed that the ML detection problem can be solved by a quantization operation. More specifically, the quantization operation is described as \cite{a9}
\begin{equation}
\hat{m}=\begin{cases}m_{1}, & {\rm if}\ \frac{||{\bf{y}}||^2}{N}< B_1; \cr m_{i}, & {\rm if}\ B_{i-1}\le \frac{||{\bf{y}}||^2}{N}\leq B_i, \ i=2, \cdots, L_m-1; \cr m_{{L_m}}, & {\rm if}\ \frac{||{\bf{y}}||^2}{N}>  B_{L_m-1}, \end{cases}
\label{messagedetector}
\end{equation}
where $B_i$ is the optimal decision threshold between $m_i$ and $m_{i+1}$. The threshold $B_i$ can be represented by
\begin{equation}
B_i=\frac{A_{i}A_{i+1}\ln{\frac{A_{i+1}}{A_{i}}}}{A_{i+1}-A_{i}},\quad i=1,...,L_m-1,
\label{solve_b}
\end{equation}
where $A_i=|m_i|^2+\sigma^2$. Based on this optimal decision rule, the correct detection probability of $i$-th symbol $m_i$, denoted by $P_{c,i}$, is determined by \cite{a9}
\begin{equation}
P_{c,i}=\begin{cases}G\left(\frac{NB_i}{A_{i}}\right), & {\rm if}\ i=1; \cr G\left(\frac{NB_i}{A_{i}}\right)-G\left(\frac{NB_{i-1}}{A_{i}}\right), &  {\rm if}\ i=2,\cdots,L_m-1; \cr 1-G\left(\frac{NB_{L_m-1}}{A_{i}}\right), & {\rm if}\ i=L_m,
\end{cases}
\end{equation}
where $G(z)$ is the cumulative distribution function (CDF) of a complex Chi-squared distribution variable $Z$, given by
\begin{equation}\label{CDF}
G(z)=1-e^{-z}\sum_{L=0}^{N-1}\frac{z^L}{L!},\ \ z>0.
\end{equation}
When each message symbol is selected from $\mathcal{M}$ with equal probability, the average message symbol error rate (SER), denoted by $P_{e}$, can be calculated as follows:
\begin{equation}
\label{SERofMessage}
P_{e}=1-\frac{1}{L_m}\sum\limits_{i=1}^{L_m}P_{c,i}.
\end{equation}
By minimizing $P_{e}$ under the constraint that average message power is not greater than $E_m$, the asymptotically optimal non-negative PAM constellation design for massive SIMO systems can be represented as follows \cite{a9}:
\begin{equation}
\left\{0, \sigma^2({R}-1),\sigma^2({R}^2-1), \cdots, \sigma^2({R}^{L_m-1}-1)\right\},
\label{constellation}
\end{equation}
where  ${R}$ (${R}>1$) is obtained by solving the equation 
\begin{equation}
\sum\limits_{j=0}^{L_m-1}{R}^j=L_m\left(\frac{E_m}{\sigma^2}+1\right).
\label{solve_r}
\end{equation}
The message signal-to-noise ratio (SNR), denoted by $\gamma_m$, is defined as ${\gamma_m}=\frac{E_m}{\sigma^2}$. The constellation points are described from the perspective of power since there is a specific correspondence between power and amplitude in non-negative PAM. Take $L_m=4$ as an example, the message constellation points described with $A_i$ and the corresponding decision thresholds $B_i$ are shown in Fig.~\ref{Constellation_a}.
\subsection{Tag Embedding Design}\label{Embedded Structure}
We are now ready to elaborate the proposed tag embedding scheme and the corresponding detection rule. One-bit embedding is considered in this paper, i.e., $L_t=2$. When the $i$-th message symbol is transmitted, the power of the embedded symbol $E_{i,j}$ has two possible values according to two different tag bits,
\begin{equation}  
E_{i,j}=\begin{cases}|m_{i}|^2+|{t_{i,1}}|^2, & {\rm if}\ j=1\ ({\rm{tag\ bit\ is\ 0)}}; \cr |m_{i}|^2+|{t_{i,2}}|^2, &{\rm if}\ j=2\ ({\rm{tag\ bit\ is\ 1)}}. \end{cases}
\end{equation}
Note that the power of tag signal ${t_{i,1}}$ and ${t_{i,2}}$ are variables depending on message signal $m_{i}$, which is the key idea of the proposed ``Message-based Tag Modulation''. In existing methods that use uniform tag embedding, ${t_{i,1}}$ and ${t_{i,2}}$ are constants for different message signal $m_{i}$. The reason we do not use uniform tag embedding is that the message constellation points in our method contain the relationship of geometric series as shown in Fig.~\ref{Constellation_a}. If ${t_{i,1}}$ and ${t_{i,2}}$ are high power tag signals, they may be suitable for signal points $A_3$ and $A_4$, but will not work well for $A_1$ and $A_2$, and vice versa. Due to the effect of the embedding tag, the message decision thresholds \eqref{solve_b} need to be updated. We also take $L_m=4$ as an example to show the tag embedding design and the corresponding decision thresholds in Fig.~\ref{Constellation}. Since the message constellation points that are close to each other are more error-prone, we use the nearest two constellation points to calculate the new message decision threshold $B_i'$ by \eqref{solve_b},
\begin{equation}
B_i'=\frac{A_{i,2}A_{i+1,1}\ln{\frac{A_{i+1,1}}{A_{i,2}}}}{A_{i+1,1}-A_{i,2}},\ i=1,\cdots,L_m-1,
\label{solve_newb}
\end{equation}
where $A_{i,j}=E_{i,j}+\sigma^2$. Using \eqref{messagedetector} and \eqref{solve_newb}, the message symbol can be estimated, which is the first step of the detection. Based on the result of estimated message symbol, tag symbol is detected subsequently with the following rule,
\begin{equation}
\hat{t}=\begin{cases}t_{i,1} ({\rm{tag\ bit\ is\ 0)}}, & {\rm if}\ \frac{||{\bf{y}}||^2}{N}\leq C_i; \cr t_{i,2} ({\rm{tag\ bit\ is\ 1)}}, & {\rm if}\ \frac{||{\bf{y}}||^2}{N}> C_i;  \end{cases}
\label{tagdetector}
\end{equation}  
where $C_i$ is the optimal decision threshold to decide which tag bit is embedded in message signal $m_i$. Since \eqref{solve_b} also follows the general form of non-coherent ML decision threshold, $C_i$ can be expressed as 
\begin{equation}
C_i=\frac{A_{i,1}A_{i,2}\ln{\frac{A_{i,2}}{A_{i,1}}}}{A_{i,2}-A_{i,1}},\ i=1, 2, \cdots, L_m.
\label{solve_c}
\end{equation}

Note that the instantaneous CSI is not required in message detector \eqref{messagedetector} and tag detector \eqref{tagdetector}. As such, successive interference cancellation (SIC) detection widely used in existing PLA methods is no longer applicable in this method. According to the constellation design results of Section \ref{pre}, the optimal power of first constellation point is zero. Therefore, we can set the power of the tag signal to zero when tag bit is $0$, i.e., $|t_{i,1}|^2=0$ (for $i=1,\cdots, L_m$).
\section{Error Performance Analysis and Optimization}\label{Opti}
In this section, the message SER and tag SER are analyzed first as the performance metrics of the proposed PLA method. Then the optimization problem of signal design is formulated and solved according to the specific system requirements.
\subsection{Error Performance Analysis}

According to the assumptions of $\bf{h}$ and $\bf{n}$ in Section \ref{SystemModelSec}, $\bf{y}$ is also a circularly symmetric complex Gaussian random vector. The mean vector of $\bf{y}$ is a zero vector, and its covariance matrix can be written as
\vspace{-1em}

\begin{small}
\begin{equation}  
\begin{split}
\mathbb{E}\left[{\bf{y}}{\bf{y}}^H\right] &=\mathbb{E}\left[\left({\bf{h}}\sqrt{|m|^2+|t|^2}+{\bf{n}}\right)\left({\bf{h}}\sqrt{|m|^2+|t|^2}+{\bf{n}}\right)^H\right] \\
&=\left(|m|^2+|t|^2+\sigma^2\right){\bf I}_N.
\end{split}
\end{equation}
\end{small}Define a new random variable $Z'$
\begin{equation}
Z'=\frac{||{\bf{y}}||^2}{|m|^2+|t|^2+\sigma^2},
\end{equation}
which follows complex Chi-squared distribution.
When $m_1$ is transmitted with embedding tag $t_{1,1}$, according to \eqref{messagedetector} and \eqref{solve_newb}, the message signal can be correctly detected if $\frac{||{\bf{y}}||^2}{N}< B_1'$. This condition is equivalent to
\begin{equation}
\frac{||{\bf{y}}||^2}{|m_1|^2+|t_{1,1}|^2+\sigma^2}< \frac{NB_1'}{|m_1|^2+|t_{1,1}|^2+\sigma^2}=\frac{NB_1'}{A_{1,1}}.
\end{equation}
Let $P_{cm,i}$ denote the average correct message detection probability of $x_i$. Then $P_{cm,1}$ can be derived as
\begin{equation}
P_{cm,1}=G\left(\frac{NB_1'}{A_{1,1}}\right).
\end{equation}
Similarly, $P_{cm,i}$ can be given by
\begin{small}
\begin{equation}
P_{cm,i}=\begin{cases}\frac{1}{2}\!\sum\limits_{j=1}^{2}G\left(\frac{NB_i'}{A_{i,j}}\right), & {\rm if}\ i\!=\!1; \cr\frac{1}{2}\!\sum\limits_{j=1}^{2}\left[G\left(\frac{NB_i'}{A_{i,j}}\right)\!-\!G\left(\frac{NB_{i-1}'}{A_{i,j}}\right)\right], &  {\rm if}\ i\!=\!2,\cdots,L_m\!-1\!; \cr \frac{1}{2}\!\sum\limits_{j=1}^{2}\left[1\!-\!G\left(\frac{NB_{L_m-1}'}{A_{i,j}}\right)\right], & {\rm if}\ i\!=\!L_m.
\label{realPcm}
\end{cases}
\end{equation}
\end{small}The average message SER $P_{em}$ can be calculated as
\begin{small}
\begin{equation}
\label{RealmessageSER}
P_{em}=1-\frac{1}{L_m}\sum\limits_{i=1}^{L_m}P_{cm,i}.
\end{equation}
\end{small}The assumption is made that the message symbols are selected from constellation points collection with equal probability. Due to the uniformity characteristic of hash functions, different tag bits can also be considered appearing with equal probability\cite{a14}. 

We now consider the error performance of the tag signal. Suitable hash functions exhibit ``avalanche effect", i.e., the output changes significantly if the input alters slightly\cite{a14}. The authentication fails when the message estimation has only one bit error, which indicates that the recalculated MAC will be meaningless if the message parts have errors. As a result, we consider the tag SER under the condition that the message symbol is detected correctly. It should be noted that the message SER can be controlled to be low enough (e.g. less than $10^{-5}$) during the embedding design process. Therefore, in this paper, the tag correct rate and error rate are calculated from the perspective of conditional probability. When the embedded symbol $x_i$ is transmitted, the tag correct detection rate under the condition that the message symbol $m_i$ is correctly obtained can be determined by \eqref{tagdetector}, which is 

\begin{small}
\begin{equation}
P_{ct,i}=\frac{1}{2}\left[G\left(\frac{NC_i}{A_{i,1}}\right)+1-G\left(\frac{NC_i}{A_{i,2}}\right)\right].
\end{equation}
\end{small}

To simplify the expression, we define the proportional variables $r_i(i=1,\cdots,L_m)$

\begin{small}
\begin{equation}
r_i=\frac{A_{i,2}}{A_{i,1}},\ 1<r_i<R.
\end{equation}  
\end{small}Therefore, the average tag SER, denoted by $P_{et}$, can be represented by
\vspace{-0.5 em}
\begin{small}
\begin{equation}
\begin{split}
P_{et} &=\frac{1}{L_m}\sum \limits_{i=1}^{L_m}\left(1-P_{ct,i}\right) \cr
           &=\frac{1}{2L_m}\sum \limits_{i=1}^{L_m}\left[1+G\left(Nu(r_i)\right)-G\left(Nv(r_i)\right)\right],
\label{Pet}
\end{split}
\end{equation}
\end{small}where $u(r_i)=\frac{\ln r_i}{r_i-1}$ and $v(r_i)=\frac{r_i\ln r_i}{r_i-1}$.
\subsection{Trade-off Characterization}\label{optimal}
One primary purpose of this paper is to design an optimal embedding scheme to minimize the tag SER subject to the total average power constraint. Meanwhile, the system reliability should meet the certain requirement, i.e., $P_{em}<\delta$, where $\delta$ is the message SER requirement threshold. This optimization problem is formulated as follows
\begin{subequations}\label{generaloptimal}
\begin{align}
&\min \limits_{\{ A_{i,j}\}_{i=1,\cdots,L_m}^{j=1,2}} \quad P_{et}\\ \label{general1}
&{\rm {s.t.}}\quad
E_t+E_m\leq E_{tot} , \\ \label{general2}
&\ \ \ \ \ \ \ P_{em} \le \delta.
\end{align}
\end{subequations}
This problem is non-trivial because of multiple optimization variables and the complex structure of constraint functions. To simplify this problem, we use two steps to solve it. First, we find the optimal tag embedding scheme when the message constellation is fixed (i.e., $E_m$ is given). Second, we search for the optimal allocated power of message $E_m$ that can minimize the tag SER. 

Now we consider the first step. When $E_m$ is fixed, the message SER $P_{em}$ and the tag SER $P_{et}$ are given by \eqref{RealmessageSER} and \eqref{Pet}, respectively. Note that $P_{em}$ is complicated since it contains multiple variables ($A_{i,j}$, $i=1,\cdots,L_m,\ j=1\ {\rm or}\ 2$). To reduce the complexity of constraint \eqref{general2}, an upper bound of message SER, $P_{em}^u$, can be derived to replace $P_{em}$ in \eqref{general2}.  The upper bound can be derived as follows
\vspace{-1.5 em}

\begin{small}
\begin{equation}
P_{em}^u=\frac{1}{L_m}\sum \limits_{i=1}^{L_m-1}\left\{1-G\left[Ng(r_i)\right]+G\left[Nh(r_i)\right]\right\},
\label{upperbound}
\end{equation}
\end{small}where $g(r_i)=\frac{{R}\ln{\frac{{R}}{r_i}}}{{R}-r_i}$, $h(r_i)=\frac{r_i\ln{\frac{{R}}{r_i}}}{{R}-r_i}$. The proof is provided in Appendix \ref{ProofUpperBoundSec}.

Using variable substitution $r_i=e^{k_i}(0<k_i<\ln{R})$ and the message SER upper bound in \eqref{upperbound}, the optimization problem \eqref{generaloptimal} can be rewritten as
\vspace{-1 em}

\begin{small}
\begin{subequations}\label{convex}
\begin{align}\label{convex1}
&\min \limits_{\{ k_i\}_{i=1}^{L_m}} \quad \frac{1}{2L_m}\sum \limits_{i=1}^{L_m}\left\{1+G\left[Nu(e^{k_i})\right]-G\left[Nv(e^{k_i})\right]\right\}\\ \label{convex2}
&{\rm {s.t.}}\quad
\frac{1}{2L_m}\sum \limits_{i=1}^{L_m}A_{i,1}(e^{k_i}-1)\leq E_{tot}-E_m, \\ \label{convex3}
&\ \ \ \ \ \ \ \frac{1}{L_m}\sum \limits_{i=1}^{L_m-1}\left\{1-G\left[Ng(e^{k_i})\right]+G\left[Nh(e^{k_i})\right]\right\} \le \delta.
\end{align}
\end{subequations}
\end{small}We can show that \eqref{convex} is a convex optimization problem and the proof is provided in Appendix \ref{ConvexProofSec}. Therefore \eqref{convex} can be efficiently solved by interior-point method and the optimal tag embedding scheme can be determined when $E_m$ is given.

In the second step, we consider the situation when $E_m$ is a variable, different $E_m$ results in different values of $P_{et}$. In this case, the optimization result of \eqref{convex} is a function of $E_m$, which is denoted by $H(E_m)$. Note that when $H(E_m)$ achieves its minimal value, the inequality \eqref{general1} will become a equality, i.e., $E_m+E_t= E_{tot} $. The reason is that if $E_{tot}$ still has a surplus, $P_{em}$ and $P_{et}$ can be further reduced by increasing message and tag power at the same time. Therefore, this optimization problem can be narrated as finding the optimal power allocation between message signal and tag signal when the constraints are still satisfied. Define the power allocation factor $\alpha$,
\begin{equation}
\alpha=\frac{E_m}{E_{tot}},\ \alpha_0\leq \alpha \leq1,
\end{equation}
where $\alpha_0$ is the minimum factor that makes $E_m$ satisfy the constraint $P_{em}^u \le \delta$.
The function $H(E_m)$ can be represented by $H(\alpha)$, which indicates that different $\alpha$ corresponds to different minimized tag SER. Therefore, the new power allocation problem can be formulated as 
\vspace{-1 em}

\begin{small}
\begin{subequations}
\begin{align}
\label{power}
&\min \limits_{\alpha_0 \leq \alpha\leq 1}\quad H(\alpha)\\ \label{power1}
&{\rm {s.t.}}\quad
E_t\leq (1-\alpha)E_{tot} , \\ \label{power2}
&\ \ \ \ \ \ \ P_{e,m}^u \le \delta.
\end{align}
\end{subequations}
\end{small}This is a problem of single variable with a limited range, which can be efficiently solved by one-dimensional search.

When we set different values of message SER requirement threshold $\delta$ for the considered system, different minimized tag SER can be obtained by solving the optimization problem \eqref{generaloptimal}. Therefore, we can characterize the trade-off between the message and tag error performance through changing the value of the message SER requirement thresholds. 
\section{Numerical Results}\label{SimulationResults}
We carried out computer simulations to verify the theoretical results. In our simulations, we set $N=128$ and $\sigma^2=1$.

\begin{figure}[bp]
\vspace{-1.0em}
\centerline{\includegraphics[width=0.45\textwidth]{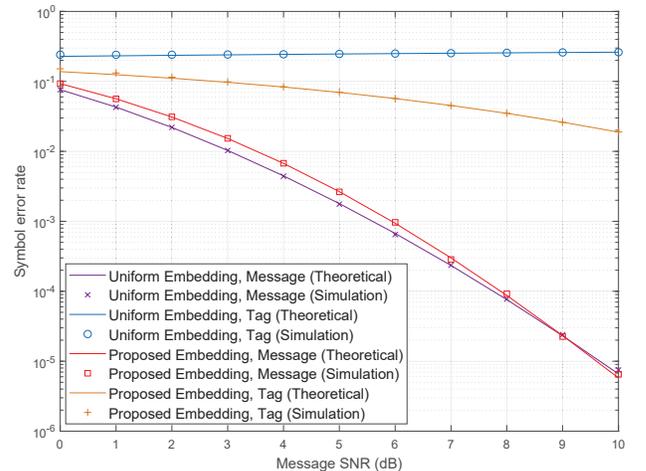}}
\vspace{-1em}
\caption{Comparison of SER of the proposed embedding scheme and uniform embedding scheme.}
\label{Fig1}
\end{figure}

Firstly, to show the accuracy of the message and tag error performance analysis, the theoretical and simulation results of $P_{em}$ and $P_{et}$ are demonstrated in Fig.~\ref{Fig1} and they match with each other. 
$E_{tot}$ is set to be large enough because optimization has not yet been considered. The ``Message-based Tag Modulation" scheme is compared with the ``Uniform Embedding" scheme that embeds same tag power levels for all message symbols. For a fair comparison, we control the tag power of two embedding schemes to make sure they have the same message SER performance (i.e., less than $10^{-5}$) when the message SNR is $10$dB. The results show that the tag SER of the proposed embedding scheme decreases as the message SNR increases. However, the tag SER of the ``Uniform Embedding" scheme is always above $90\%$. Therefore, ``Uniform Embedding" scheme is not suitable for PLA in non-coherent massive SIMO communications.

\begin{figure}[tbp]
\centerline{\includegraphics[width=0.45\textwidth]{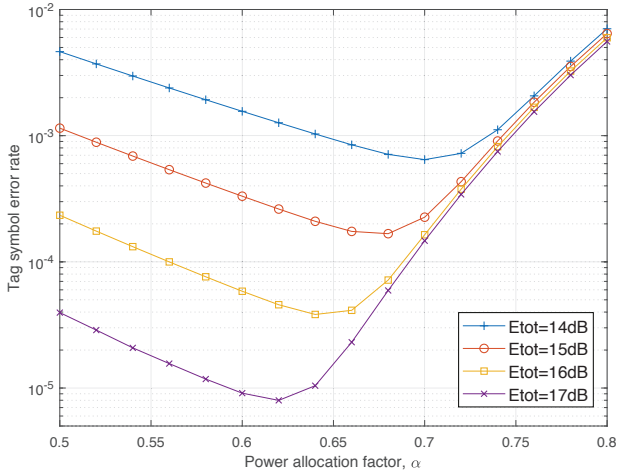}}
\vspace{-1em}
\caption{Optimization results of tag SER with different $E_{tot}$.}
\vspace{-1.0em}
\label{Fig2}
\end{figure}

Then we focus on the optimal performance by solving optimization problems in Section \ref{optimal}. The message SER upper bound threshold $\delta$ is set to $10^{-5}$. The results of optimization problem \eqref{convex} and the search results for power allocation are provided in Fig.~\ref{Fig2}. The tag SER decreases as the total system average power increases and the optimal power allocation factor can be found from the results.

\begin{figure}[tbp]
\centerline{\includegraphics[width=0.45\textwidth]{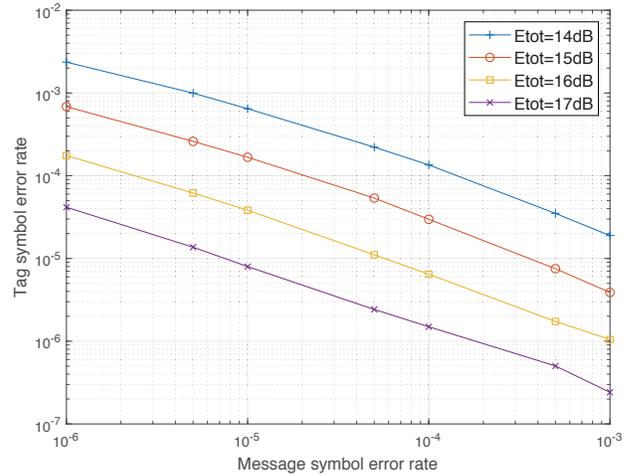}}
\vspace{-1em}
\caption{Trade-off curves of tag SER and message SER with different $E_{tot}$.}
\vspace{-1.0em}
\label{Fig3}
\end{figure}

Finally, to understand the entire system performance, the trade-off between $P_{em}$ and $P_{et}$ is depicted in Fig.~\ref{Fig3}. We can observe from the trade-off curves that the tag SER decreases as the message SER requirement threshold increases. The results in Fig.~\ref{Fig3} also show that both tag SER and message SER can be reduced when the total system average power increases. The trade-off curve presents the optimal tag SER performance under different system requirements for message SER, which can provide useful insights for practical PLA system design.
\section{Conclusions}\label{Conclusion}
In this paper, we proposed an active PLA mechanism for non-coherent massive SIMO-based IIoT systems. This paper is the first to show that active PLA can be achieved without the need of pilot signal and channel estimation. We designed the optimal tag embedding scheme when the message constellation is given. Then we solved the power allocation problem to obtain the optimal tag SER performance. The trade-off curve between tag SER and message SER was depicted to offer a comprehensive understanding of the system performance. From the simulation results, we can conclude that ``Message-based Tag Modulation" is necessary for the considered non-coherent system. Moreover, the proposed authentication method can meet the specific power and error rate requirements of IIoT system. As an initial effort, the tag embedding design was limited to 1-bit tag per message symbol in this paper. We will extend our work to multiple bits tag per message symbol for future work.  
\appendices
\section{Proof of an Upper Bound of the Message SER}\label{ProofUpperBoundSec}
Notice that the average message SER of $x_1$ is less than transmitting $\sqrt{{m_1}^2+{t_{1,2}}^2}$, because close constellation points produce a large message SER. Let $P_{em,i}$ denote the average message SER when $m_i$ is transmitted. $P_{em,1}$ and its upper bound can be derived as follows
\vspace{-1em}

\begin{small}
	\begin{equation}
	\begin{aligned}
	P_{em,1}&=\frac{1}{2}\Bigg[{\rm P}\left(\hat{m}=m_2,\cdots,m_{L_m}|x_1=\sqrt{{m_1}^2+{t_{1,1}}^2}\right) \cr 
	&\ \ \  +{\rm P}\left(\hat{m}=m_2,\cdots,m_{L_m}|x_1=\sqrt{{m_1}^2+{t_{1,2}}^2}\right)\Bigg] \cr
	&<{\rm P}\left(\hat{m}=m_2,\cdots,m_{L_m}|x_1=\sqrt{{m_1}^2+{t_{1,2}}^2}\right) \cr
	&=1-G\left(\frac{NB_1'}{A_{1,2}}\right).
	\end{aligned}
	\end{equation}
\end{small}Similarly, the average message SER of $x_{L_m}$ is less than transmitting $\sqrt{{m_{L_m}}^2+{t_{L_m,1}}^2}$, so the upper bound of $P_{em,L_m}$ can be derived as follows
\vspace{-1em}

\begin{small}
	\begin{equation}
	\begin{aligned}
	P_{em,L_m}&<{\rm P}\left(\hat{m}=m_1,\cdots,m_{L_m-1}|x_{L_m}=\sqrt{{m_{L_m}}^2+{t_{L_m,1}}^2}\right) \cr
	&=G\left(\frac{NB_{L_m-1}'}{A_{Lm,1}}\right).
	\end{aligned}
	\end{equation}
\end{small}Further, the average message SER of $x_i( i=2, \cdots, L_m-1)$ also has an upper bound due to the same reason above,
\vspace{-1em}

\begin{small}
	\begin{equation}
	\begin{aligned}
	P_{em,i}&<{\rm P}\left(\hat{m}=m_1,\cdots,m_{i-1}|x_i=\sqrt{{m_i}^2+{t_{i,1}}^2}\right) \cr 
	&\ \ \ +{\rm P}\left(\hat{m}=m_{i+1},\cdots,m_{L_m}|x_i=\sqrt{{m_i}^2+{t_{i,2}}^2}\right)\cr
	&=1-G\left(\frac{NB_i'}{A_{i,2}}\right)+G\left(\frac{NB_{i-1}'}{A_{i,1}}\right).
	\end{aligned}
	\end{equation}
\end{small}From the above, an upper bound of $P_{em}$ can be derived as

\begin{small}
	\begin{equation}
	\begin{aligned}
	P_{em} &=\frac{1}{L_m}\sum \limits_{i=1}^{L_m}P_{em,i} \cr
	& < \frac{1}{L_m}\sum \limits_{i=1}^{L_m-1}\left[1-G\left(\frac{NB_i'}{A_{i,2}}\right)+G\left(\frac{NB_i'}{A_{i+1,1}}\right)\right] \cr 
	& =\frac{1}{L_m}\sum \limits_{i=1}^{L_m-1}\left\{1-G\left[Ng(r_i)\right]+G\left[Nh(r_i)\right]\right\},
	\end{aligned}
	\end{equation}  
\end{small}where $g(r_i)=\frac{{R}\ln{\frac{{R}}{r_i}}}{{R}-r_i}$, $h(r_i)=\frac{r_i\ln{\frac{{R}}{r_i}}}{{R}-r_i}$. This completes the proof of an upper bound of the message SER.

\section{Proof of the convex optimization problem}\label{ConvexProofSec}
The considered optimization problem is rewritten as follows:
\vspace{-1 em}

\begin{small}
	\begin{subequations}\label{convex}
		\begin{align}\label{convex1}
		&\min \limits_{\{ k_i\}_{i=1}^{L_m}} \quad \frac{1}{2L_m}\sum \limits_{i=1}^{L_m}\left\{1+G\left[Nu(e^{k_i})\right]-G\left[Nv(e^{k_i})\right]\right\}\\ \label{convex2}
		&{\rm {s.t.}}\quad
		\frac{1}{2L_m}\sum \limits_{i=1}^{L_m}A_{i,1}(e^{k_i}-1)\leq E_{tot}-E_m, \\ \label{convex3}
		&\ \ \ \ \ \ \ \frac{1}{L_m}\sum \limits_{i=1}^{L_m-1}\left\{1-G\left[Ng(e^{k_i})\right]+G\left[Nh(e^{k_i})\right]\right\} \le \delta.
		\end{align}
	\end{subequations}
\end{small}

Let $F(k)=1+G\left[Nu(e^{k})\right]-G\left[Nv(e^{k})\right]$ and $W(k)=1-G\left[Ng(e^{k})\right]+G\left[Nh(e^{k})\right]$. 

$F(k)$ is a convex function for $0<k<\ln{R}$ according to the lemma in \cite{a9}. The objective function in \eqref{convex1} is a sum of $L_m$ convex functions $F(k_i)( i=1,\cdots,L_m)$ and its Hessian matrix is a diagonal matrix, thus the Hessian matrix is positive definite and the objective function in \eqref{convex1} is a convex function. Note that $e^k-1$ is a basic convex function. The left side of \eqref{convex2} is a sum of $L_m$ convex functions $e^{k_i}-1(i=1,\cdots,L_m)$ and its Hessian matrix is a diagonal matrix, so it is also a convex function. The left side of \eqref{convex3} is a sum of $L_m-1$ functions $W(k_i)( i=1,\cdots,L_m-1)$ and its Hessian matrix is a diagonal matrix. Therefore, the optimization problem \eqref{convex} is a convex problem if $W(k)$ is a convex function for $0<k<\ln{R}$.

Let $W_1(k)=G\left[Ng(e^{k})\right]$ and $W_2(k)=G\left[Nh(e^{k})\right]$. Then we have $W'(k)=W'_2(k)-W'_1(k)$. The derivative of $G(z)$ is $f_Z(z)=\frac{1}{(N-1)!}z^{N-1}e^{-z}$. Then we have
\vspace{-1em}

\begin{small}
	\begin{equation}
	\begin{aligned}
	W'_1(k)&=f_Z\left[Ng(e^k)\right]Ng'(e^k) \cr
	&=\frac{1}{(N-1)!}\left[Ng(e^k)\right]^{N-1}e^{-Ng(e^k)}Ng'(e^k).
	\end{aligned}
	\end{equation}
\end{small}Since $h(e^k)=g(e^k)\frac{e^k}{R}=g(e^k)+(k-\ln{R})$, we can obtain that $h'(e^k)=\frac{1}{R}e^k\left[g(e^k)+g'(e^k)\right]$. Then we have
\vspace{-1em}

\begin{small}
	\begin{equation}
	\begin{aligned}
	W'_2(k)&=f_Z\left[Nh(e^k)\right]Nh'(e^k) \cr
	&=\frac{1}{(N-1)!}\left[Nh(e^k)\right]^{N-1}\cdot e^{-Nh(e^k)}Nh'(e^k) \cr
	&=\frac{1}{(N-1)!}\left[Ng(e^k)\frac{e^k}{R}\right]^{N-1}\cdot e^{-N\left[g(e^k)+(k-\ln{R})\right]} \cr &\ \ \  \cdot N\left\{\frac{1}{R}e^k\left[g(e^k)+g'(e^k)\right]\right\} \cr
	&=W'_1(k)+\frac{1}{(N-1)!}N^Ng(e^k)^{N}e^{-Ng(e^k)}.
	\end{aligned}
	\end{equation}
\end{small}Therefore, we can simplify the $W'(k)$ as follows

\begin{small}
	\begin{equation}
	W'(k)=\frac{1}{(N-1)!}N^Ng(e^k)^{N}e^{-Ng(e^k)}.
	\end{equation}
\end{small}The second-order derivative of $W(k)$ can be obtained by
\vspace{-1em}

\begin{small}
	\begin{equation}
	W''(k)=\frac{N^{N+1}g(e^k)^{N-1}e^{-Ng(e^k)}g'(e^k)\left[1-g(e^k)\right]}{(N-1)!},
	\end{equation}
\end{small}where 
\vspace{-1em}

\begin{small}
	\begin{equation}
	\begin{aligned}
	g'(e^k)&=\frac{R\left[e^k-R+e^k(\ln{R}-k)\right]}{(R-e^k)^2} \cr
	&\overset{r=e^k}{=}\frac{Rr}{(R-r)^2}(1+\ln{\frac{R}{r}}-\frac{R}{r}).
	\end{aligned}
	\end{equation}
\end{small}Since $\ln{x}-x$ is a monotonically decreasing function for $x>1$, then we have $1+\ln{\frac{R}{r}}-\frac{R}{r}<0$. Therefore, $g'(e^k)<0$ for $0<k<\ln{R}$. Then we know that $g(e^k)$ is a decreasing function when $0<k<\ln{R}$. Moreover, $\lim_{k \to \ln{R}}g(e^k)=1$, then $1-g(e^k)<0$ for $0<k<\ln{R}$. We can conclude that $W''(k)>0(0<k<\ln{R})$, thus $W(k)$ is a convex function for $0<k<\ln{R}$. Above all, the optimization problem \eqref{convex} is a convex problem.

\end{document}